\documentstyle[prl,epsf,amsbsy,amstext,amsfonts,amssymb,aps]{revtex}

\begin{document}

\title{Exclusive electromagnetic production of strangeness on the 
nucleon~: review of recent data in a Regge approach.}
\author{M. Guidal $^{a}$, J.-M. Laget $^{b}$ and M. Vanderhaeghen$^{c}$}
\address{$^{a}$ IPN Orsay, F-91406 Orsay, France}
\address{$^{b}$ CEA/Saclay, DAPNIA/SPhN, F-91191 Gif-sur-Yvette Cedex,
France}
\address{$^{c}$ University Mainz, D-55099 Mainz, Germany}
\date{\today}
\maketitle

\begin{abstract}
In view of the numerous experimental results recently released, 
we provide in this letter an update on the performance of our simple 
Regge model for strangeness electroproduction on the nucleon. 
Without refitting any parameters, a decent 
description of all measured observables and channels is achieved. 
We also give predictions for spin transfer observables, 
recently measured at Jefferson Lab which have high sensitivity
to discriminate between different theoretical approaches. 
\end{abstract}


%\twocolumn[\vspace*{12.cm}]
\twocolumn

Regge theory provides a simple and elegant framework to describe 
exclusive hadronic reactions above the resonance 
region~\cite{regge1,regge2,storrow}. We have presented in 
references~\cite{ourpapers1,ourpapers2,ourpapers3,ourpapers4})
a Regge model for meson photo- and electroproduction reactions
which is based on the exchange of one or two
meson Regge trajectories in the $t$-channel. The very few
free parameters in this approach are the coupling constants of
the first particle materialization of the Regge trajectory at the
hadronic vertices 
%(i.e. $g_{KYN}, g_{K*YN}, \kappa_{K^*YN}$), 
along with the mass scales 
%($\Lambda_K$ and $\Lambda_{K*}$) 
of monopole form factors at the electromagnetic vertices in the case of 
electroproduction.
Basically, all available observables are decently and ``economically" 
described by this approach for a plethora of elementary channels~: 
photo- and electroproduction on the nucleon of $\pi^{0,\pm}$ and $K^+$ as well as
$\rho^0, \omega, \phi,\gamma$~\cite{laget1,laget2} and 
$\eta, \eta^\prime$~\cite{taiwan}.
Surprisingly, in some cases, data down to W$<2$ GeV center of mass energies,
therefore supposedly in the resonance region, can be succesfully described~:
this could be interpreted as a manifestation of the reggeon-resonance
duality hypothesis~\cite{dolen}. Why this duality seems to work for some channels 
and not for some others still remains an open question.

Recently, numerous experimental data have been released by the 
Jefferson Lab, ELSA, GRAAL and SPRING8 facilities in the
kaon production sector. This motivates this study which compares the Regge 
model to these new observables and channels, without any change of the parameters. 
Our model is fully described in Refs.~\cite{ourpapers1,ourpapers2,ourpapers3,ourpapers4}. 
In those works, we found that, for strangeness electromagnetic production, it allows
to describe the $\gamma^{(*)} + p \rightarrow K^+ 
+ \Lambda$ and $\gamma^{(*)} + p \rightarrow K^+ + \Sigma$ reactions through
the exchange of only two trajectories in the $t$-channel~: $K$ and $K^*$.
The coupling constants at the ($K,(\Lambda,\Sigma),N$) and
($K^*,(\Lambda,\Sigma),N$) vertices were derived and fitted from the 
photoproduction study~\cite{ourpapers1,ourpapers2} where all existing 
high energy data could be satisfactorily described~:
\begin{eqnarray}
&&g_{K\Lambda N}=-11.54 \; , \; g_{K^*\Lambda N}=-23.0 \; , 
\; \kappa_{K^*\Lambda N}=2.5, \nonumber\\
&&g_{K\Sigma N}=4.48 \; \; \; , \;  g_{K^*\Sigma N}=-25.0 \; ,
\; \kappa_{K^*\Sigma N}=-1.0
\label{ctes}
\end{eqnarray}

In the electroproduction case, the other parameters of the model
are the two (squared) mass scales of the monopole form factors at the 
$\gamma,K,(K,K^*)$ vertices, which were taken both equal to 1.5 
GeV$^2$~\cite{ourpapers4} in order to fit the high $Q^2$ behavior of 
the data. 

Finally, let's recall that one essential feature of the model is the way gauge
invariance is restored for the $t$-channel $K$ exchange by proper
``reggeization" of the $s$-channel nucleon pole contribution. As detailed
further below, this is the key element to describe the slow decrease with $Q^2$
of the $R=\sigma_L /\sigma_T$ ratio for the $K^+ \Lambda$ 
channel, a feature which was found
to be difficult to accomodate in all other approaches.

\begin{figure}[h]
\epsfxsize=9.5 cm
\epsfysize=10. cm
%\centerline{\epsffile[20 72 543 216]{mohring.ps}}
\centerline{\epsffile{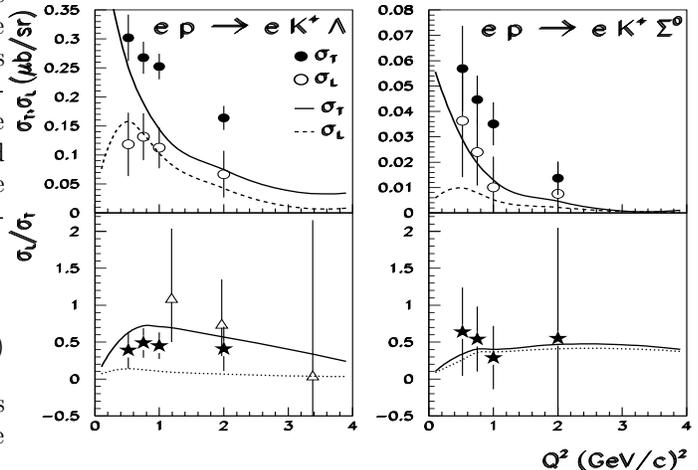}}
\vspace{-1.cm}
\caption[]{$Q^2$ dependence of $\sigma_T$ and $\sigma_L$ (upper panels) and
of their ratio $R=\sigma_L /\sigma_T$ (lower panels) for the 
$\gamma^* + p \rightarrow K^+ + \Lambda$ reaction (left panels) and
$\gamma^* + p \rightarrow K^+ + \Sigma$ reaction (right panels), at 
W = 1.84 GeV. Experimental data points are from~:
($\bullet$, $\circ$, $\bigstar$)~: Ref.~\cite{mohring} and 
($\triangle$)~: Ref.~\cite{bebek77b}.
For the lower panels, the full curve is the model with reggeized
$K$ and $K^*$ $t$-channel exchanges whereas the dotted curve has standard 
Feynman poles for the $K$ and $K^*$ propagators.}
\label{fig:mohring}
\end{figure}

These data, first published in~\cite{baker}, have actually been 
recently re-analysed~\cite{mohring} by the JLab collaboration E93018. 
We compare on Fig.~\ref{fig:mohring} the Regge model predictions~\cite{ourpapers4} 
with these new $Q^2$ dependence for the experimental transverse ($\sigma_T$) and 
longitudinal ($\sigma_L$) cross sections, along with their ratio ($R$), 
for both the $\Lambda$ and $\Sigma$ channels. 
The corrections due to the new analysis are non-negligible for
the absolute values of the longitudinal and transverse 
cross sections and affect significantly the slopes of the $Q^2$ dependences.
Our (unchanged) model gives now a much better description of $\sigma_L$ but
it significantly underestimates $\sigma_T$ at large $Q^2$, for both 
channels. However, the ratio $R$ is still very well reproduced and displays
a slow decrease as $Q^2$ increases. 

The curves of the lower panels of Fig.~\ref{fig:mohring} illustrate that 
the origin of the decrease with $Q^2$ of the ratio $R$ is actually not 
so much the result of the Regge treatment of the $K$ and $K^*$ $t$-channel 
propagators but, rather, as mentionned before, of the particular way gauge 
invariance is restored (i.e. by reggeizing the nucleon $s$-channel and kaon 
$t$-channel diagrams and assigning to them the {\it same} electromagnetic form 
factor). The reggeization is nevertheless necessary to ensure a correct 
normalization of ($\sigma_T$) and ($\sigma_L$) as standard Feynman propagators 
would produce cross sections higher by factors of 2 to 5 at these energies.

\vspace{-.5 cm}

A better agreement with these new data could certainly be achieved by changing
the values of the $K$ and $K^*$ form factor mass scales but this would destroy
the nice agreement with the other kaon electroproduction 
data~\cite{bebek77b,brauel79,feller,bebek77,azemoon,brown,bebek74}, 
for which $W>$ 2.1 GeV and which were presented in Ref.~\cite{ourpapers4}. We 
prefer to interpret this discrepancy as room for potential additionnal 
processes at these lower energies (W=1.84 GeV), such as $s$-channel
resonances. It has indeed already been observed that the well known nucleon resonances
have larger transverse photo-couplings than longitudinal ones~\cite{pdg,burkert}.

\begin{figure}[h]
\epsfxsize=9.5 cm
\epsfysize=10. cm
%\centerline{\epsffile[20 72 543 216]{mohring2.ps}}
\centerline{\epsffile{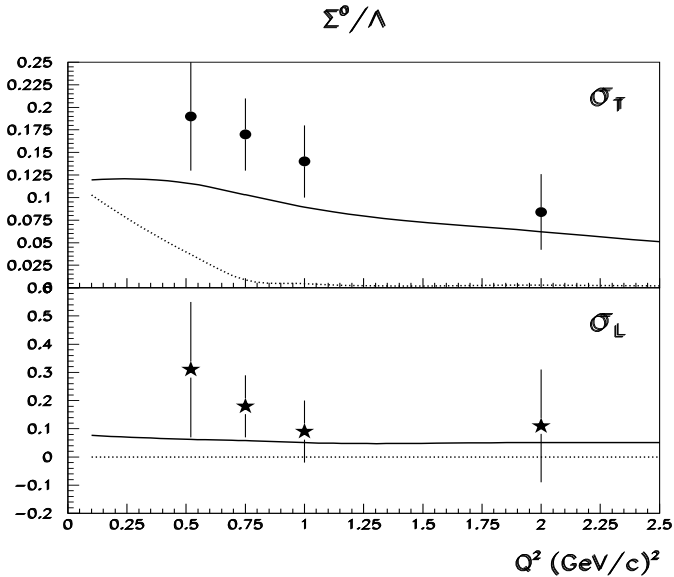}}
\vspace{-1.cm}
\caption[]{$Q^2$ dependence of the ratio of the $\gamma^* + p 
\rightarrow K^+ + \Sigma$ to the $\gamma^* + p \rightarrow K^+ + \Lambda$ 
forward ($\theta_K^{cm}$=0) differential cross sections at 
W = 1.84 GeV. Upper panel: transverse cross section; lower panel: 
longitudinal cross section. Full curve~: $K$+$K^*$ exchanges; dotted
curve~: only $K^*$ exchange. Experimental data points are 
from Ref.~\cite{mohring}.}
\label{fig:mohring2}
\end{figure}

Fig.~\ref{fig:mohring2} shows the $Q^2$ dependence of the $\Sigma \over
\Lambda$ ratio for both transverse and longitudinal cross sections. Again,
a good agreement with the data is found without any additional
refitting of the parameters of the model. In our $t$-channel approach,
only the kaon exchange contributes to $\sigma_L$ (like for the pion form 
factor in pion electroproduction, the best way to access the kaon form 
factor at intermediate and large $Q^2$ is to isolate $\sigma_L$, as it is 
mostly insensitive to $K^*$ exchange).
This is why the $\Sigma\over\Lambda$ ratio for $\sigma_L$ is basically
constant in our model. However, for the transverse part of the cross section,
the $K$ and $K^*$ contribute with different weights for the $\Lambda$ and 
$\Sigma^0$ channels and therefore, the $\Sigma\over\Lambda$ ratio is no
longer constant for $\sigma_T$.

\begin{figure}[h]
\epsfxsize=9.5 cm
\epsfysize=10. cm
%\centerline{\epsffile[20 72 503 216]{schuh2.ps}}
\centerline{\epsffile{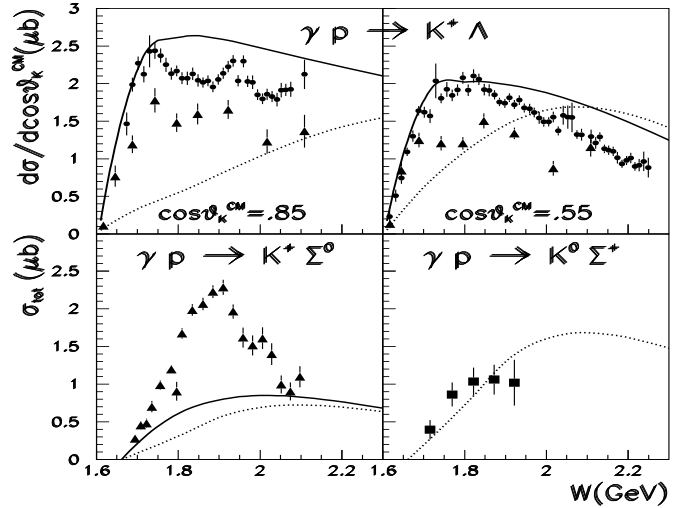}}
\vspace{-1.5cm}
\caption[]{$W$ dependence of the $\gamma + p \rightarrow K^+ + \Lambda$
{\it differential} cross section at cos($\theta_K^{CM}$) $\approx$ 0.85  
and cos($\theta_K^{CM}$) $\approx$ 0.55 (left and right upper panels
respectively) and of the $\gamma + p \rightarrow K^+ + \Sigma$ and 
$\gamma + p \rightarrow K^0 + \Sigma^+$ {\it total} cross sections (left
and right lower panels respectively). Full curve~: $K$+$K^*$ exchanges; dotted
curve~: only $K^*$ exchange. Experimental data points are 
from Refs.~\cite{schuh} (circles), ~\cite{bonn} (triangles) 
and~\cite{bonn2} (squares).}
\label{fig:schuh}
\end{figure}

Fig.~\ref{fig:schuh} displays the photoproduction data with, in particular, the
latest results from the JLab/CLAS collaboration~\cite{schuh} which are
about 20-25\% above the previous Bonn/Saphir data~\cite{bonn}. For the $\Lambda$ 
channel, there are structures in the experimental data (``bumps" around 
W$\approx$ 1.75 GeV and 1.95 GeV) which hint to $s$-channel resonances 
excitations and which our model can obviously not reproduce
as it is purely a $t$-channel model ; however, the JLab experimental data 
(for the forward angles) lies now, in average, very close to the Regge model,
unchanged from Ref.~\cite{ourpapers1,ourpapers2}. 
This supports, at the 10-15\% level and for this channel, the concept of a global 
duality between the sum of all $t$-channel exchanges and all $s$-channels excitations.
 
\vspace{.5cm}

As to the $\Sigma^0$ channel, the Bonn/Saphir data~\cite{bonn} largely
exceeds the calculation of our model. The data point to a prominent 
$s$-channel resonance structure around W $\approx$ 1.9 GeV. It would be interesting
to explore the kinematical region $W>$ 2.1 GeV to see if our model overestimates
the experimental data which would therefore make up for the underestimation of our
model in the $W<$ 2.1 GeV and which would restore, in average, the duality idea.
This high energy domain can be explored by the JLab CLAS collaboration up
to $E_\gamma \approx$ 6 GeV (W$\approx$ 3.5 GeV).

The $K^0 \Sigma^+$ channel can be calculated in a straightforward way,
without any additional  parameter by taking into account 
only the $K^{0*}$ $t$-channel exchange (as a real photon cannot elastically
couple to the neutral spinless $K^0$) and using \(g^2_{\gamma K^0 K^{*0}}=
g^2_{\gamma K^+ K^{*+}} \Gamma_{K^{*0}\to\gamma K^0} / {\Gamma_{K^{*+}\to\gamma K^+}} 
\approx\) 2.3  \(g^2_{\gamma K^+ K^{*+}}\) with the radiative widths taken from 
Ref.~\cite{pdg}. Figure~\ref{fig:schuh} shows a very nice agreement between 
our calculation and the Bonn/Saphir data of Ref.~\cite{bonn2} which can be interpreted as 
an absence of $s$-channel excitation in this channel (which brings constraint
on the isospin of the hinted resonance around W$\approx$1.9 GeV decaying into 
the $\Sigma^0$ channel). If the $\Sigma^+$ channel
is indeed produced by pure $t$-channel exchange, these data provide
us with a strong constraint (and in our case, a nice confirmation) of 
the strength of the $K^*$ exchange, which is the only allowed leading Regge 
trajectory and whose coupling constants have been derived 
independently~\cite{ourpapers2}.

\vspace{.5cm}
\begin{figure}[h]
\epsfxsize=9.5 cm
\epsfysize=10. cm
%\centerline{\epsffile[20 72 503 216]{spring8.ps}}
\centerline{\epsffile{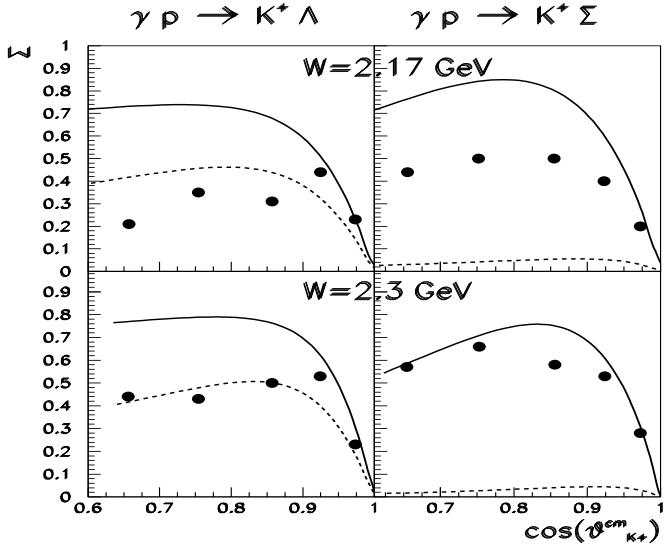}}
\vspace{-1.5cm}
\caption[]{cos($\theta_{K^+}^{cm}$) dependence of the photon asymmetry 
for the $\gamma + p \rightarrow K^+ + \Lambda$ (left)
and $\gamma^* + p \rightarrow K^+ + \Sigma$ (right) processes at 
W = 2.17 GeV (or $E_\gamma$=2.05 GeV) and W = 2.3 GeV (or $E_\gamma$=2.35 GeV). 
Experimental data points are from Ref.~\cite{spring8}. Full curve~: $K$+$K^*$
exchanges, dashed curve~: $K$ exchange.}
\label{fig:spring8}
\end{figure}

Turning now to polarization observables, Fig.~\ref{fig:spring8} shows the 
photon asymmetry recently measured by the Spring-8 collaboration at
the LEPS facility~\cite{spring8} in photoproduction for both the $\Lambda$ and
$\Sigma$ channels. The general trend of the data is reproduced by the model, in
particular the sharp rise of the beam asymmetry at very forward angles (indication
of the dominance of the natural parity K* exchange).
At smaller center of mass energies and larger angles, the discrepancies
between theory and data become more pronounced. This should not come
as a surprise as Regge theory is essentially valid at high energies and forward
angles. It is nevertheless certainly of great interest to probe the limits
of this model through such data.
Let's also remind that the Regge model describes well at forward angles
the magnitude and the sign of the $\Lambda$ and $\Sigma$ recoil 
polarizations~\cite{ourpapers4}.

\begin{figure}[h]
\epsfxsize=9.5 cm
\epsfysize=10. cm
%\centerline{\epsffile[20 72 503 216]{carman2.ps}}
\centerline{\epsffile{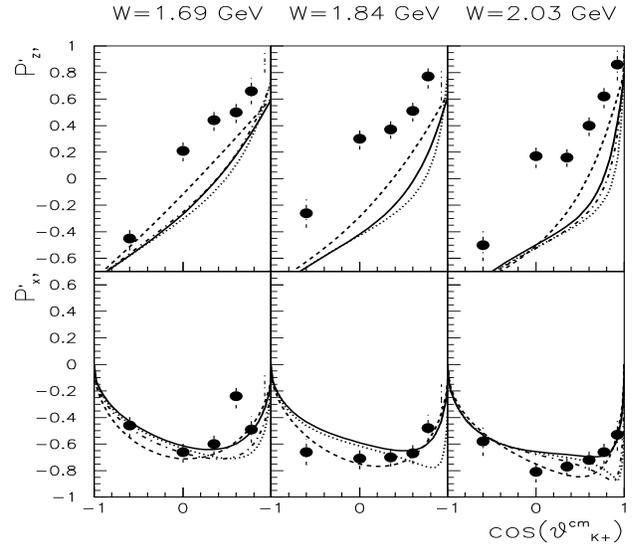}}
\vspace{-1.5cm}
\caption[]{cos$\theta_{K^+}^{cm}$ dependence of the transfered $\Lambda$ 
polarization in $e + p \rightarrow e^\prime + K^+ + \Lambda$ (left)
for $E_e$ = 2.5 GeV and $<Q^2>$= 0.8 GeV$^2$. Full curve~: $K$+$K^*$ reggeized exchange.
Dashed curve~: $K$ reggeized exchange. Dotted curve~: $K^*$ reggeized exchange.
Dash-dotted curve~: $K$+$K^*$ standard Feynman pole exchange. Experimental 
data points are from Ref.~\cite{carman}.}
\label{fig:carman1}
\end{figure}

An electroproduction experiment at JLab~\cite{carman} has measured 
for the first time, the transfered polarisation from a longitudinally 
polarized beam to the $\Lambda$ recoil hyperon in the exclusive reaction
$e + p \rightarrow e^\prime + K^+ + \Lambda$. 
We show on Fig.~\ref{fig:carman1}
the only two transferred polarization components $P^\prime_{x^\prime}$ and
$P^\prime_{z^\prime}$\footnote{where $x^\prime$ and $y^\prime$ refer
to the cartesian coordinates in the $\gamma^*-p$ center of mass frame
with the $z$-axis along the direction of the produced kaon} which are non-zero 
when integrated over $\Phi$, the azimuthal angle between the leptonic and the 
hadronic planes.
The longitudinal spin transfer component ($P^\prime_{z^\prime}$) is well 
reproduced at the very forward angles. The sideways component spin 
transfer component ($P^\prime_{x^\prime}$) is well reproduced over
the full angular range.

\vspace{.5cm}

We have plotted, along with the standard $K$+$K^*$ reggeized exchanges
which constitutes our full model, the individual contributions of the
reggeized $K$ and $K^*$ exchanges. We also show the calculation
carried out with standard Feynman propagators of the type 
$1/(t-m^2_{(K,K^*)})$, instead of Regge propagators of the type
$s^{\alpha_{(K,K^*)}(t)}$. It can be seen that these observables are actually
barely sensitive to these variations and that basically any model 
based on $K$ and/or $K^*$ $t$-channel exchange, whatever is the chosen prescription 
(Feynman or Regge type pole), gives a decent description of the data. However,
it should be noted that, when introducing $s$-channel resonances processes
(see, for instance the isobaric models cited in Ref.~\cite{carman}), such double
polarization observables are very sensitive to resonance properties
and allow to discriminate rather precisely between various sets of nucleon 
resonances participating in the reaction. 

In summary, the latest experimental results released in the domain of open strangeness 
electromagnetic production on the nucleon confirm that our ``simple" Regge model surprisingly 
reproduces the gross features of the data, even for $W<$ 2 GeV. It thus
provides an economical description and a simple explanation of the data,
hinting that a sort of reggeon-resonance duality is at work here. Where it fails, 
it gives a useful hint that other mechanisms than simple $t$-channel mechanisms are necessary.
In the difficult task of putting into evidence new nucleon resonances in the
strangeness channel which are overlapping
or hidden into large backgrounds, this sort of clue, i.e. the contribution of 
non-resonant $t$-channel and Born mechanisms, is certainly much needed. 

This work was supported in part by the French CNRS/IN2P3, the French
Commissariat \`a l'Energie Atomique, the Deutsche Forschungsgemeinschaft 
(SFB443) and the European Commission IHP program 
(contract HPRN-CT-2000-00130).

\end{document}